\documentclass[a4paper]{article}
\newdimen\footheight
\usepackage{sprocl}
\usepackage{epsfig}

\newcommand{\LCDM}{\bf \Lambda{\rm \bf CDM }}

\def\be{\begin{equation}}
\def\ee{\end{equation}}
\def\bea{\begin{eqnarray}}
\def\eea{\end{eqnarray}}

\begin{document}

\title{PROPERTIES OF GALAXY CLUSTERS}

\author{ALEXANDER KNEBE}

\address{Astrophysikalisches Institut Potsdam, Germany}

\maketitle\abstracts{
The formation process, properties, and spatial distribution of galaxy groups and 
clusters are closely related to the background cosmological model. We use 
numerical simulations of variants of the CDM model with different cosmological 
parameters in order to compare
statistical measures such as mass spectra, merging histories, correlation 
functions, and clustering power spectra with relevant observations. 
The study of the time evolution shows that the 
internal structure of galaxy groups and clusters is connected with their 
merging history. Here we discuss the virialisztion and the halo's spin 
parameters.}

\section{Cosmological N-body simulations}

The initial realization at redshift z=25 is achieved by 
generating a glass distribution and then displacing the 
particles according to the Zel'dovich approximation.
The evolution of this gravitating system of dark matter particles is
simulated with an extended version of the adaptive 
${\rm P}^3$M code (originally developed by H.P.Couchman) which is able 
to allow for other background expansions than the Einstein-de-Sitter 
universe. The simulations, which are all COBE-normalized, are run with $128^3$ 
particles on a $128^3$ grid with a force resolution of 50 ${\rm h}^{-1}$kpc.
Their other (physical) parameters are as follows:

\medskip

\begin{center}
\begin{tabular}{|l||c|c|c|} \hline
 {\bf Simulation}  & {\bf SCDM}  & {$\LCDM$}  & {\bf OCDM}   \\ \hline
 Box [${\rm h}^{-1}{\rm Mpc}$]  & 200 & 280 & 280         \\ \hline
 $\Omega_0$, $\Omega_{\Lambda}$
                                   & 1.0, 0.0 
                                         & 0.3, 0.7 
                                               & 0.5, 0.0    \\ \hline
  H$_0$ [km/sec/Mpc]
              & 50.0      & 70.0      & 70.0       \\ \hline
 $\sigma_8$   & 1.18      & 1.00      & 0.96       \\ \hline
\end{tabular}
\end{center}

\section{Identification of galaxy clusters}

The identification of galaxy clusters is performed with a 
friends-of-friends algorithm using a dimensionless linking 
length of 0.2. In order to avoid misidentifications we check 
the virial theorem 
$|E_{\rm pot}| = 2 \ E_{\rm kin}$ for each individual halo. 

Following Spitzer (1969), the potential energy
can be approximated by $ |E_{\rm~pot,approx}|~\approx~0.4~G~M^2~/~r_h$ 
where $r_h$ is the half mass radius of the system. 
Scatter plots of $E_{\rm pot}$ against $E_{\rm kin}$ show the expected 
linear behavior for most of the clusters. However,
some of them have much too high velocity dispersions 
compared to their potential energy. These clusters may be smaller
objects connected by a linking bridge, random encounters of particles,
or unbound halos. Therefore, they are treated separately in the further 
analysis and investigated in detail.\cite{1}

Figure 1 (left panel) shows the virial theorem for the P$^3$M simulation in
comparison with a PM simulation run both using the $\Lambda$CDM model 
with exactly the same starting realization. The PM dynamics underestimates
the inner halo velocities. The 'lower' virial relation found within the 
AP$^3$M simulation can be understood as an influence of an outer pressure 
of radially infalling particles into the halos.\cite{2}

\begin{figure}
\psfig{figure=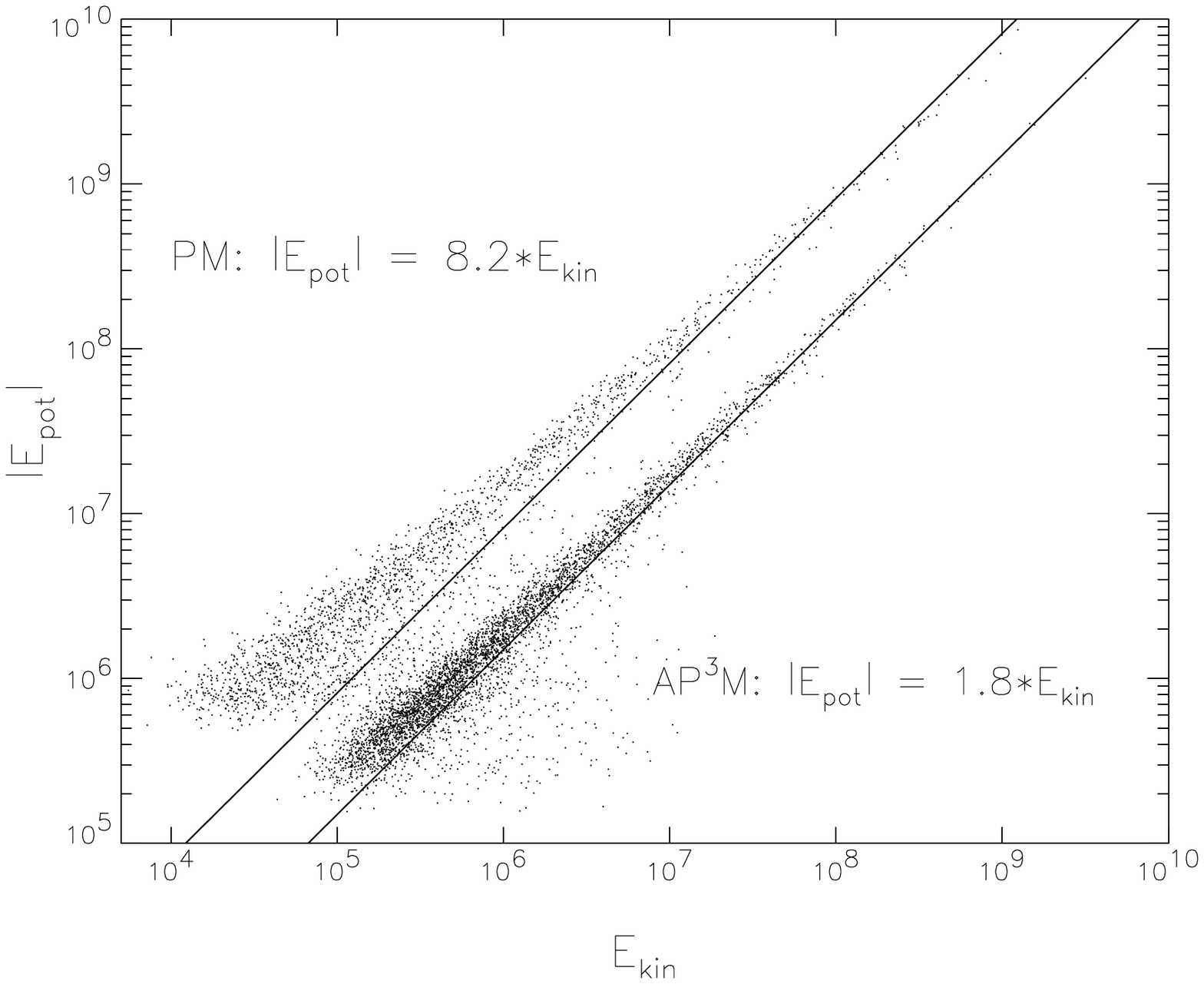,  height=0.27\textheight}
\psfig{figure=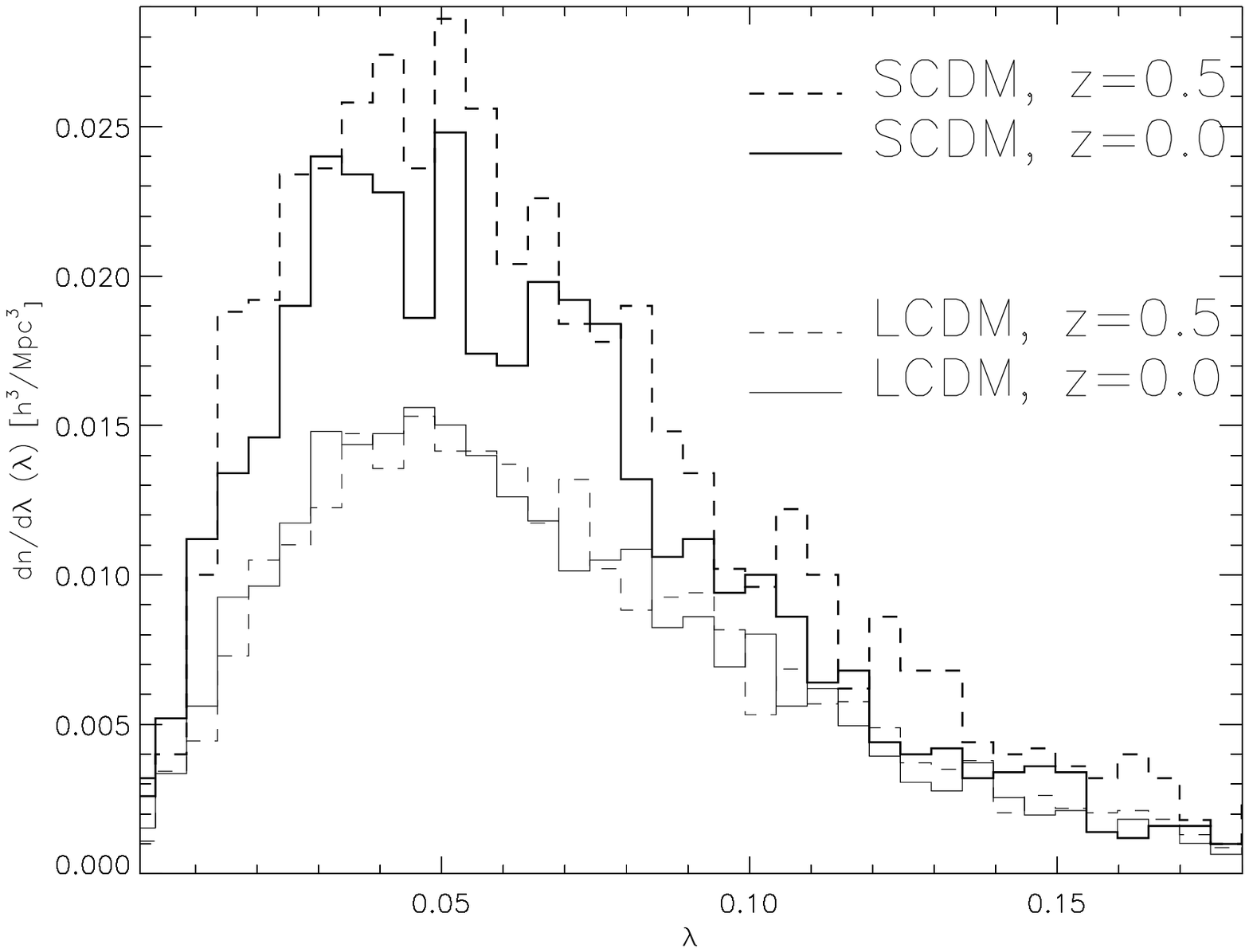,  height=0.27\textheight}
\caption{Virial theorem for clusters (left) and 
         spin parameter distributions (right)}
\end{figure}

\section{Spin parameter}

The spin parameter $\lambda := J \sqrt{|E|} / (G M^{5/2})$ for each bound 
halo is calculated and used to extract the distribution 
$dn (\lambda) / d\lambda$, i.e. the number of clusters with spin parameter 
$\lambda$ per unit volume in the interval $[\lambda, \lambda+d\lambda]$.

Figure 1 (right panel) shows $f(\lambda)$ for the SCDM and the 
$\Lambda$CDM model. The graph for the OCDM model looks almost identical
to the $\Lambda$CDM model and is left out for clarity.
There is no significant evolution detectable within the presented 
redshift interval from z=0.0 to z=0.5. Angular momentum grows mainly during
the early quasilinear stages where there the different models approximately
coincide.\cite{3}

\end{document}